\documentclass[a4paper,12pt,dvips,]{scrartcl}
\usepackage{graphicx}

\newcommand\beq{\begin{equation}}
\newcommand\eeq{\end{equation}}
\newcommand\bea{\begin{eqnarray}}
\newcommand\eea{\end{eqnarray}}

\begin{document}
\begin{titlepage} 
\vspace{0.2in}

\begin{center} {\LARGE \bf 
Mixmaster Chaoticity as  
Semiclassical Limit of the 
Canonical Quantum Dynamics   
} 
\vspace*{2cm}

{\bf Giovanni Imponente$~^{1,3,4}$  \\
Giovanni Montani$~^{2,4}$ } 

\vspace*{2cm}
$~^1$ Dipartimento di Fisica Universit\'a di Napoli ``Federico II'' \\
$~^2$ Dipartimento di Fisica Universit\'a di Roma ``La Sapienza'' \\
$~^3$ INFN --- Napoli \\ 
$~^4$ ICRA --- International Center for Relativistic Astrophysics \\ 
c/o Dip. Fisica (G9), Piazzale Aldo Moro 5, 00185 Roma, Italy\\
\vspace*{1cm}
e-mail: imponente@icra.it, montani@icra.it

\vspace{\stretch{1}}

\end{center} \indent


\begin{abstract}

Within a cosmological framework, we provide a 
Hamiltonian analysis of the Mixmaster Universe dynamics 
on the base of a standard Arnowitt-Deser-Misner approach,
showing how the chaotic behavior characterizing the evolution 
of the system near the cosmological singularity can be obtained
as the semiclassical limit of the canonical quantization of the model
in the same dynamical representation. \\
The relation between this intrinsic chaotic behavior 
and the indeterministic quantum dynamics 
is inferred through the coincidence
between the microcanonical probability distribution 
and the semiclassical quantum one.
\end{abstract}

\end{titlepage}

\section{Introduction}

The simplest and most interesting generalization 
of the Friedmann-Lemaitre-Robertson-Walker (FLRW) cosmology is
the Bianchi IX model, whose geometry has the
homogeneity constraint but the dynamics 
makes allowance of anisotropic 
evolution of different (linearly independent) 
spatial directions. 

Belinski, Kalatnikov and Lifshitz (BKL) at the 
end of Sixties \cite{BKL70}
derived the oscillatory regime characterizing 
the asymptotic evolution near the cosmological singularity, describing 
via a discrete map
the resulting chaos. \\
This Mixmaster \cite{M69} dynamics, whose geometry is 
invariant under the $SO(3)$ group, 
allows the line element 
to be decomposed as a FLRW model 
plus a gravitational waves packet (\cite{L74}, \cite{GDY75})
sufficiently far
from the singularity. \\

The various approaches 
in terms of continuous variables (i.e. construction 
of an invariant measure for the system \cite{CB83,KM97}
and study of the system covariance \cite{CL97}-\cite{IM02})
have shown how this chaotic feature is invariant 
with respect to any choice of the temporal gauge and how
in this sense it is intrinsic. \\
%
%
Such deep nature of the Mixmaster 
deterministic chaos and its very early appearance in the Universe
evolution lead to believe in the existence of a relation
with the quantum behavior the system performs during the 
Planckian era. \\
The aim of this paper is to give a precise meaning to this 
relation by constructing the semiclassical limit of a Sch\"oedinger
approach to the canonical quantization of the 
Arnowitt-Deser-Misner (ADM) dynamics 
whose corresponding probability distribution 
coincides with the (deterministic) microcanonical one.\\
In Section \ref{billiard} are outlined the stochastic properties 
of the Mixmaster dynamics, being isomorphic to a billiard 
on a Lobachevsky plane and is given a stationary 
statistical distribution within the framework of the 
microcanonical ensemble. \\
In Section \ref{limit} we show the existence of a direct 
correspondence between the classical and quantum dynamics
outlined by the common form of the continuity equation for 
the statistical distribution and the one for the first order 
approximation in the semiclassical expansion. \\
As a remarkable feature, both formalisms (the classical and the 
quantum one) are constructed
in a generic temporal gauge.

\section{Statistical Mechanics Approach}\label{billiard}

\begin{figure}[htbp]
\begin{center}
\includegraphics[width=8cm,clip]{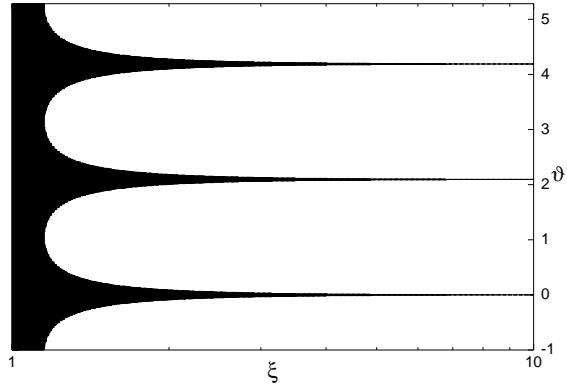} 
\end{center}
\caption{Domain $\Gamma_H$, where the 
anisotropy parameters $H_a$ are simultaneously greater than 0. \label{fig:domain}}
\end{figure}

As well known (see \cite{IM01}), the dynamics of 
the Bianchi type IX model, in terms of generic 
Misner-Chitr\`e--like (MCl) variables 
$(\tau, \xi, \theta)$ and corresponding 
coniugate momenta, is summarized by the variational
principle 
\begin{equation}
\label{qnew} 
\delta \int \left(   p_{\xi} \frac{d\xi}{d\tau} 
+  p_{\theta} \frac{d\theta}{d\tau} 
- \frac{d\Gamma}{d\tau}{\cal H}_{ADM} \right) d\tau =0 \, ,
\end{equation}
where 
\begin{equation} 
\label{n2}
{\cal H}_{ADM} = \sqrt{\varepsilon ^2 +U}\,  , \qquad
\varepsilon ^2 = \left({\xi}^2 -1\right){p_{\xi}}^2 +\frac{{p_{\theta}}^2}{{\xi}^2 -1} 
\end{equation}
and $U(\tau, \xi, \theta)$ denotes a potential term which, 
asymptotically to the singularity, is modeled by the 
potential wall 
\begin{eqnarray}
\label{aa}
U_{\infty} = 
&\Theta _\infty \left(H_l\left(\xi, \theta\right)\right) + 
\Theta _\infty \left(H_m\left(\xi, \theta\right)\right) + 
\Theta _\infty \left(H_n\left(\xi, \theta\right)\right) \, \\ 
&\Theta _\infty \left(x\right) = \biggl\{ 
\begin{array}{lll} 
+ \infty & {\rm if } & x < 0 \\ 
\quad 0 & {\rm if } & x >  0 
\end{array} \nonumber  \; ,
\end{eqnarray}
being $H_a$ ($a=l,m,n$) the anisotropy parameters \cite{IM01},
such that the motion of the point Universe is restricted
in the domain $\Gamma_H$
(see Figure \ref{fig:domain}). 
Such a dynamical scheme has the relevant feature to let 
free the choice of the lapse function unless 
explicitly chosen the form of the function $\Gamma$.
The bounces against the potential walls and the instability 
of the geodesic flow on the Lobachevsky plane let the 
dyna\-mics acquire a stochastic 
feature.
Moreover in $\Gamma_H$ 
the ADM Hamiltonian 

becomes (asymptotically) an ``energy-like'' integral of motion
$d {\cal H}_{ADM}/d \Gamma = 0$, say 
${\cal H}_{ADM}=\varepsilon = E ={\rm const.}$.

As shown in \cite{IM01,IM02}, the Jacobi metric associated 
with the principle (\ref{qnew}) 
describes a point--universe moving over 
a Lobachevsky plane reduced by the potential walls (\ref{aa}) 
to a billiard.

The Statistics of the Mixmaster stochasticity
is reformulated following the 
lines presented in \cite{KM97}, \cite{M00} and \cite{IM02}
to derive the invariant measure. \\
This system  
is well-described by a {\it microcanonical ensemble}, whose 
Liouville invariant measure $w_{\infty}$ can be obtained as a solution 
of the continuity equation
\begin{equation}
\label{cont}
\sqrt{\xi^2-1}\cos\phi \frac{\partial w_{\infty}}{\partial \xi}+
\frac{\sin\phi}{\sqrt{\xi^2-1}}\frac{\partial w_{\infty}}{\partial \theta}-
\frac{\xi\sin\phi}{\sqrt{\xi^2-1}}\frac{\partial w_{\infty}}{\partial \phi}=0 \, .
\end{equation}
Over the reduced phase space\footnote{$S^1_{\phi}$ 
denotes the $\phi$-circle of the momentum space 
variable.} 
$\{\xi,\theta\}\otimes S^1_{\phi}$,   $w_{\infty}$
explicitly reads 
\begin{equation}
\label{step}
w_{\infty}\left(\xi, \theta, \phi\right)=\left\{ \begin{array}{lll} \displaystyle
\frac{1}{8\pi^2} \qquad &\forall& \left\{ \xi, \theta, \phi\right\} \in \Gamma_H\otimes S^1_{\phi}  \\ 
0  \qquad &\forall& \left\{ \xi, \theta, \phi\right\} \not\in \Gamma_H\otimes S^1_{\phi} 
\end{array} 
\right. \, .
\end{equation}
Using generic MCl variables, 
the above invariant measure is independent 
of the choice of the 
temporal gauge, i.e. of the lapse function.

In order to restrict the invariant measure 
to the two-dimensional space $\{ \xi , \theta \}$, 
we stress how the original 
Hamiltonian equations in the asymptotic limit 
for which $U\rightarrow U_{\infty}\Rightarrow \varepsilon=E={\rm const.}$ get in 
$\Gamma_H$ the free geodesic motion \cite{M00}
 \begin{equation}
\label{ham}
\frac{d\xi}{d\tau}=\frac{d\Gamma}{d\tau}\sqrt{\xi^2-1}\cos\phi \, , \qquad 
\frac{d\theta}{d\tau}=\frac{d\Gamma}{d\tau}\frac{\sin\phi}{\sqrt{\xi^2-1}} \, ,\qquad 
\frac{d\phi}{d\tau}=-\frac{d\Gamma}{d\tau}\frac{\xi\sin\phi}{\sqrt{\xi^2-1}} \, . 
\end{equation}
By this system, we easily rewrite the stationary
continuity equation in the form
\begin{equation}
\label{con1}
\partial _{\phi }w_{\infty } +
\frac{d\xi }{d\phi }\partial _{\xi }w_{\infty } +
\frac{d\theta }{d\phi }\partial _{\theta }w_{\infty } = 0
\, .
\end{equation}
When assuming the independence of the distribution
function $w_{\infty }$ from $\phi$, then
the normalization condition leads to the restricted function 

\begin{equation}
\label{conf}
\varrho_{\infty}\left(\xi, \theta\right)\equiv
\int _0^{2\pi } w_{\infty }(\xi , \theta , \phi )d\phi = 
2\pi w_{\infty}(\xi , \theta )
\, . 
\end{equation}
Thus, we get the following continuity equation for
$\varrho _{\infty }$
\begin{equation}
\label{cont2}
\sqrt{\xi^2-1}\cos\phi \frac{\partial \varrho_{\infty}}{\partial \xi}+
\frac{\sin\phi}{\sqrt{\xi^2-1}}\frac{\partial\varrho_{\infty}}{\partial \theta}=0 \, ,
\end{equation}
where $\phi$ plays the role of a parameter and 
the corresponding {\it microcanonical} solution on the whole configuration
space $\{\xi, \theta\}$ reads 
\begin{equation}
\label{steprho}
\varrho_{\infty}\left(\xi, \theta\right)=\left\{ \begin{array}{lll} \displaystyle
\frac{1}{4\pi} \qquad &\forall& \left\{ \xi, \theta\right\} \in \Gamma_H  \\ 
0  \qquad &\forall& \left\{ \xi, \theta\right\} \not\in \Gamma_H 
\end{array} 
\right. \, . 
\end{equation}

We conclude observing how the Hamilton equations retain the
same form even when the potential walls have a finite
(non-zero) value. 

\section{Semiclassical Limit of the Quantum Dynamics}\label{limit}

What above outlined is a Mixmaster intrinsic feature and 
not an effect induced by a particular class of references: the whole
MCl formalism and its consequences has been developed in a framework 
free from the choice of a specific time gauge. \\
Since chaos 
appears close enough to the Big Bang,
we infer that it has some relations with the indeterministic 
quantum dynamics the model performs in the {\it Planckian era}. 
This relation between quantum and deterministic chaos is searched
in the sense of a semiclassical limit for a canonical 
quantization of the model.

The asymptotical principle (\ref{qnew}) describes a two dimensional 
Hamiltonian system, which can 
be quantized by a natural Schr\"oedinger approach 
\begin{equation}
\label{sch}
i \hbar \frac{\partial \psi}{\partial \tau}
=\frac{d\Gamma}{d\tau}\hat{{\cal H}}_{ADM}\psi \, ,
\end{equation}
being $\psi=\psi(\tau,\xi,\theta)$ the wave 
function for the point-universe
and, implementing $\hat{{\cal H}}_{ADM}$ 
(see (\ref{n2})) to 
an operator\footnote{The only non vanishing 
canonical commutation relations are
\[ \left[ \hat{\xi},\hat{p_{\xi}}\right] =i\hbar \, , \qquad \left[ \hat{\theta},\hat{p_{\theta}}\right] =i\hbar \, .
\]}, i.e.
\begin{eqnarray}
\label{op}
\xi &\rightarrow &\hat{\xi} \, , \qquad \qquad 
 \quad \qquad  \theta \rightarrow \hat{\theta} \, ,  \nonumber \\
p_{\xi} &\rightarrow & \hat{p_{\xi}} 
\equiv -i \hbar \frac{\partial}{\partial \xi} \, , \qquad 
p_{\theta} \rightarrow \hat{p_{\theta}} 
\equiv -i \hbar \frac{\partial}{\partial \theta} \, ,
\end{eqnarray} 
the equation (\ref{sch}) rewrites explicitly, in 
the asymptotic limit $U\rightarrow U_{\infty}$,
\begin{eqnarray}
\label{sch1}
i \frac{\partial \psi}{\partial \tau} &=& 
\frac{d\Gamma}{d\tau}\sqrt{\hat{\varepsilon}^2 
+\frac{U_{\infty}}{\hbar^2}}~\psi  = \nonumber\\
&=& \frac{d\Gamma}{d\tau}
\left[-\sqrt{\xi^2 -1}\frac{\partial }{\partial \xi} \sqrt{\xi^2 -1} \frac{\partial }{\partial \xi} -
\frac{1}{\sqrt{\xi^2 -1}}\frac{\partial }{\partial {\theta}}\frac{1}{\sqrt{\xi^2 -1}}\frac{\partial }{\partial {\theta}} +\frac{U_{\infty}}{\hbar^2}\right]^{1/2} \psi \, ,
\end{eqnarray}
where we took an appropriate symmetric 
normal ordering prescription 
and we left $U_{\infty}$ to stress that 
the potential cannot be neglected 
on the entire configuration space 
$\{\xi, \theta\}$. Being $U_{\infty}$ equal to infinity out 
of $\Gamma_H$, $\psi$ requires as boundary condition 
to vanish outside the potential walls, say
\begin{equation}
\label{bound}
\psi\left(\partial \Gamma_H\right)=0 \, .
\end{equation}
The {\it quantum} equation (\ref{sch1}) is equivalent to the 
Wheeler-DeWitt one for the same Bianchi model, once separated the positive and negative
frequencies solutions \cite{KU81}, with the advantage that now $\tau$ 
is a real time variable.
Since the potential walls $U_{\infty}$ are time independent, 
a solution of this equation has the form
\begin{equation}
\label{sol}
\psi\left(\tau, \xi, \theta\right)= \sum_{n=1}^{\infty} c_n e^{-i E_n \Gamma(\tau) / \hbar} \varphi_n\left(\xi, \theta\right)
\end{equation}
where $c_n$ are constant coefficients and we assumed a 
discrete ``energy'' spectrum because the quantum 
point-universe is restricted in the 
finite region $\Gamma_H$; the position (\ref{sol}) 
in (\ref{sch1}) leads 
to the eigenvalue problem 
\begin{eqnarray}
\label{eig}
\left[-\sqrt{\xi^2 -1}\frac{\partial }{\partial \xi} \sqrt{\xi^2 -1} \frac{\partial }{\partial \xi} -
\frac{1}{\sqrt{\xi^2 -1}}\frac{\partial }{\partial {\theta}}\frac{1}{\sqrt{\xi^2 -1}}
\frac{\partial }{\partial {\theta}} \right] \varphi_n = \nonumber \\
= {\left(\frac{{E_n}^2-U_{\infty}}{\hbar^2}\right)} \varphi_n 
\equiv \frac{{E_{\infty}}^2_n}{\hbar^2} \varphi_n     \, . 
\end{eqnarray} 
In what follows we search the semiclassical 
solution of this equation regarding the 
eigenvalue ${E_{\infty}}_n$ as a finite 
constant (i.e. we consider the potential 
walls as finite) and only at the end of 
the procedure we will take the limit for $U_{\infty}$.

We infer that, in the semiclassical limit 
when $\hbar \rightarrow 0$ and the 
``occupation number'' $n$ tends to infinity 
(but $n\hbar$ approaches a finite value),
the wave function $\varphi_n$ approaches a 
function $\varphi$ as 
\begin{equation}
\label{phi}
\mathop {\lim }\limits_{\scriptstyle n \to \infty  \hfill \atop 
\scriptstyle \hbar  \to 0 \hfill}
\varphi_n\left(\xi, \theta\right) =\varphi\left(\xi, \theta\right) \, , \qquad
\mathop {\lim }\limits_{\scriptstyle n \to \infty  \hfill \atop 
\scriptstyle \hbar  \to 0 \hfill}
{E_{\infty}}_n ={E_{\infty}} \, .
\end{equation}
The expression for $\varphi$ is taken as 
a semiclassical expansion up to the first order, i.e. 
\begin{equation}
\label{expa}
\varphi\left(\xi, \theta \right) =\sqrt{r \left(\xi, \theta \right)} 
\exp\left\{ i\frac{S\left(\xi, \theta \right)}{\hbar}\right\} \, ,
\end{equation}
where $r$ and $S$ are functions to be determined. \\
Substituting (\ref{expa}) in (\ref{eig}) and 
separating the real from the complex part, we 
get two independent equations, i.e.
\begin{eqnarray}
\label{1eq}
\displaystyle
{E_{\infty}}^2&=& \underbrace{\left( \xi^2-1\right) \left( \frac{\partial S}{\partial \xi}\right)^2 + 
\frac{1}{\xi^2-1} \left( \frac{\partial S}{\partial \theta}\right)^2 }_{\rm classical~term} +  \nonumber \\
&-& \frac{\hbar^2}{\sqrt{r}} \left[ \sqrt{\xi^2-1}\frac{\partial}{\partial \xi}\sqrt{\xi^2-1}\frac{\partial}{\partial \xi} + 
\frac{1}{\xi^2-1}\frac{\partial^2}{\partial \theta ^2} \right]\sqrt{r} \, ,   
\end{eqnarray}
where we multiplied both sides by $\hbar^2$ 
and, respectively,
\begin{equation}
\label{2eq}
\underbrace{\sqrt{\xi^2-1} \frac{\partial }{\partial \xi} \left( \sqrt{\xi^2-1}~r \frac{\partial S}{\partial \xi} \right) 
+\frac{1}{\xi^2-1}\frac{\partial }{\partial \theta} \left( r \frac{\partial S}{\partial \theta}\right)}_{O(1/\hbar)} =0 \, .
\end{equation}
In the limit $\hbar \rightarrow 0$ the second 
term of (\ref{1eq}) is negligible
meanwhile the first one reduces to the 
Hamilton-Jacobi equation 
\begin{equation}
\label{hamb}
\left( \xi^2-1\right) \left( \frac{\partial S}{\partial \xi}\right)^2 + 
\frac{1}{\xi^2-1} \left( \frac{\partial S}{\partial \theta}\right)^2={E_{\infty}}^2 \, .
\end{equation} 
The solution of (\ref{hamb}) can be easily checked to be\footnote{The discontinuity of this 
function on the boundary of $\Gamma_H$ is due to the model 
adopted and does not affect the probability distribution.}  
\begin{equation}
\label{esse}
S\left(\xi, \theta\right) = \int\left\{ \frac{1}{\sqrt{\xi^2 -1}}\sqrt{E^2_{\infty}
 - \frac{k^2}{\xi^2-1}}~d\xi +k ~d\theta\right\} \, , \qquad k={\rm const.}\, .
\end{equation}
We observe that (\ref{hamb}), through the identifications 
\begin{equation}
\label{ide}
\frac{\partial S}{\partial \xi} =p_{\xi} \, , \quad \frac{\partial S}{\partial \theta}=p_{\theta} 
\quad\Longleftrightarrow \quad S=\int\left( p_{\xi} d\xi + p_{\theta} d \theta \right)    \, ,
\end{equation} 
is reduced to the algebraic relation
\begin{equation}
\label{hamr}
\left( \xi^2-1\right) {p_{\xi}}^2 + \frac{1}{\xi^2-1} {p_{\theta}}^2={E_{\infty}}^2 \, .
\end{equation}
The constraint (\ref{hamr}) is nothing more than the asymptotic one 
${\cal H}_{ADM}^2=E^2={\rm const.}$ and can be solved by setting 
\begin{equation}
\label{sethm}
\frac{\partial S}{\partial \xi} =p_{\xi} \equiv \frac{E_{\infty}}{\sqrt{\xi^2-1}}\cos \phi \, , \qquad 
\frac{\partial S}{\partial \theta}=p_{\theta}\equiv E_{\infty}\sqrt{\xi^2-1}\sin \phi \, ,
\end{equation}
where $\phi\in [0,2\pi[$ is a momentum-function related 
to $\xi$ and $\theta$ by the dynamics.
On the other hand, by (\ref{esse}) we get 
\begin{eqnarray}
\label{pix}
p_{\xi} &=& \frac{1}{\sqrt{\xi^2 -1}}\sqrt{E^2_{\infty}
 - \frac{k^2}{\xi^2-1}} \\
\label{pth}
p_{\theta} &=&k \, ;
\end{eqnarray}
to verify the compatibility of these expressions with (\ref{sethm})
we use the equations of motion (\ref{ham})

\footnote{Indeed, when
the ``energy'' of the system is greater than the
finite potential walls, the classical motion is a free one.}
\begin{equation}
\label{eqmo}
\frac{d\xi}{d\phi}= - \frac{\xi^2-1}{\xi}{\rm ctg}\phi \Rightarrow 
\sqrt{\xi^2-1}\sin\phi =c \, , \qquad c={\rm const.} \, .
\end{equation}
The required compatibility comes from the identification $k=E_{\infty}c$.
Since 
\begin{equation}
\label{lim}
\mathop {\lim }\limits_{\scriptstyle U \to U_{\infty}  \hfill \atop}
E_{\infty} = \left\{ 
\begin{array}{ll}
E \qquad \, \, \, \forall  \left\{ \xi, \theta \right\} \in \Gamma_H \\
i \infty \qquad  \forall \left\{ \xi, \theta \right\} \not\in \Gamma_H 
\end{array} \right.
\end{equation} 
we see by (\ref{esse}) that the solution $\varphi\left(\xi, \theta\right)$ vanishes, 
as due in presence of infinite potential walls, outside $\Gamma_H$. \\
The substitution in (\ref{2eq}) of the positions 
(\ref{sethm})
(when the potential walls become infinite, 
because of chaotic dynamics,               
$\phi$ behaves like an independent variable and plays
here the role of a free parameter)
leads to the new equation
\begin{equation}
\label{cont3}
\sqrt{\xi^2-1}\cos\phi \frac{\partial r}{\partial \xi}+
\frac{\sin\phi}{\sqrt{\xi^2-1}}\frac{\partial r}{\partial \theta}=0 \, .
\end{equation}
We emphasize how this equation coincides with (\ref{cont2}), 
provided the identification $r\equiv\varrho_{\infty}$; it is just 
this correspondence between the statistical and the 
semiclassical quantum analysis to ensure that the indeterminism of
the quantum dynamics for the Bianchi IX model approaches the 
deterministic chaos in the considered limit. \\
Any constant function is a solution of (\ref{cont3}), but the 
normalization condition requires $r=1/4\pi$ and therefore 
we finally get 
\begin{equation}
\label{lim2}
\mathop {\lim }\limits_{\scriptstyle n \to \infty  \hfill \atop 
\scriptstyle \hbar  \to 0 \hfill}
\mid \varphi_n \mid^2  = \mid \varphi \mid^2 \equiv \varrho_{\infty} =
\left\{ \begin{array}{lll} \displaystyle
\frac{1}{4\pi} \qquad &\forall& \left\{ \xi, \theta\right\} \in \Gamma_H  \\ 
0  \qquad &\forall& \left\{ \xi, \theta\right\} \not\in \Gamma_H 
\end{array} 
\right. \, ,
\end{equation} 
say the limit for the quantum probability 
distribution as $n\rightarrow \infty$
and $\hbar \rightarrow 0$ associated to 
the wave function 
\begin{equation}
\label{swf}
\psi\left(\tau, \theta, \xi\right) 
= \varphi \left(\xi, \theta\right) 
~{\rm exp}\left\{-i \frac{E_{\infty}}{\hbar}\Gamma(\tau)\right\} 
=\sqrt{r} ~{\rm exp} \left\{i  \int \left(p_{\xi}d\xi + p_{\theta}d\theta - E_{\infty} d\Gamma \right) \right\}
\end{equation}
coincides with the classical statistical distribution on the microcanonical ensemble. \\
Though this formalism of correspondence remains valid for all Bianchi models,
only the types VIII and IX admit a normalizable wave function $\varphi(\xi, \theta)$, 
being confined in $\Gamma_H$, and a continuity equation (\ref{cont2}) which 
has a real statistical meaning. \\
Since referred to stationary states $\varphi_n(\xi, \theta)$, the considered 
semiclassical limit has to be intended in view of a ``macroscopic'' one and is not
related to the temporal evolution of the model \cite{KM97a}.



\small
\begin{thebibliography}{99} 

\bibitem{BKL70} 
Belinski V A, Khalatnikov I M and Lifshitz E M 
(1970) {\it Adv.\ Phys.},{\bf 19}, 525. 

\bibitem{M69} 
Misner C W  
(1969) {\it  Phys. Rev. Lett.} ,{\bf 22}, 1071.


\bibitem{L74}
Lukash V N, 
(1974) {\it Sov. Phys.-JETP}, {\bf 40}, n. 5, 792-799.


\bibitem{GDY75}
Grischchuk L P, Doroshkevich A G and Yudin V M 
(1975) {\it Sov. Phys.-JETP}, {\bf 42}, n.6, 943-949.

\bibitem{CB83} 
Chernoff D F and Barrow J D 
(1983) {\it Phys. Rev. Lett}, {\bf 50}, 134. 

\bibitem{KM97} 
Kirillov A A and Montani G
(1997) {\it Phys. Rev. D}, {\bf 56}, n. 10, 6225. 

\bibitem{CL97} 
Cornish N J and Levin JJ
(1997) {\it Phys. Rev. Lett.}, {\bf 78}, 998; (1997) {\it Phys. Rev. D}, {\bf 55}, 7489. 


\bibitem{M00} 
Montani G
(2000) in: {\it The Chaotic Universe, Gurzadyan V G and Ruffini R (eds)},  173, World Sci. 


\bibitem{ML00}
Motter A E and Letelier P S
avaliable gr-qc/0011001.

\bibitem{IM01}
Imponente G P, Montani G
(2001) {\it Phys. Rev. D}, {\bf 63}, 103501.

\bibitem{IM02}
Imponente G P, Montani G
(2002) to appear on {\it Int. Journ. Mod. Phys. } 
(available gr-qc/0106028).

\bibitem{KU81}
Kuchar K 
(1981) in {Quantum Gravity 2, a Second Oxford Symposium}, edited by C.J.Isham, R.Penrose and D.W.Sciama, Clarendon Press Oxford, 1981, 229.

\bibitem{KM97a}
Kirillov A A and Montani G 
(1997) {\it JETP Lett.}, {\bf 66}, 7, 475.


\end{thebibliography}
\end{document}